\documentclass[a4paper,11pt]{article}
\usepackage{pos}

\usepackage{bm}

\def \<{\langle}
\def \>{\rangle}
\def \+{\dagger}
\def \({\left(}
\def \){\right)}

\def \[{\left[}
\def \]{\right]}

\def \vp {\bm{p}}

\def \vx {\bm{x}}
\def \vv {\bm{v}}
\def \vp {\bm{p}}
\def \vk {\bm{k}}

\def \d {\delta}
\def \e {\epsilon}

\def \l {\lambda}

\def \hydro{{\rm hydro}}

\def \peak {{\rm eff}} 
\def \eff {{\rm eff}} 

\title{Beyond the wake: non-hydrodynamic response of an expanding Quark-gluon plasma}
 \ShortTitle{Beyond the wake}

\author[a,b]{Weiyao Ke}

\author*[c,d]{Yi Yin}

\affiliation[a]{Department of Physics, University of California, Berkeley\\
366 LeConte Hall, Berkeley CA 94720, United States}

\affiliation[b]{Nuclear Science Division, MS 70R0309\\
Lawrence-Berkeley National Laboratory\\
1 Cyclotron Rd, Berkeley CA 94720, United States}

\affiliation[c]{Quark Matter Research Center, Institute of Modern Physics\\
, Chinese Academy of Sciences, Lanzhou, Gansu, 073000, China}

\affiliation[d]{University of Chinese Academy of Sciences\\
Beijing, 100049, China}

\emailAdd{WeiyaoKe@lbl.gov}
\emailAdd{yiyin@impcas.ac.cn}

\abstract{
We study the response function which describes the evolution of energy density induced by an initial disturbance for a Bjorken-expanding quark-gluon plasma (QGP). 
We compare the results from solving linearized Boltzmann equation under the relaxation time approximation with those from viscous hydrodynamics.
While in long time limit the response becomes hydrodynamics, the non-hydrodynamic response is important at intermediate times when excitations of wavelength larger than the inverse of the relaxation time $1/\tau_{R}$ are not fully damped. 
Therefore observables sensitive to the jet-induced medium excitations might be employed to explore the properties of QGP at the ``mesoscopic scale''.
}

\FullConference{%
  HardProbes2020\\
  1-6 June 2020\\
  Austin, Texas}

\begin{document}
\maketitle

\section{The ``mesoscopic'' physics of quark-gluon plasma}

%

The behavior of quark-gluon plasma (QGP) at wavelength longer than its inverse temperature $1/T$  (and consequently longer than its Debye screening length) is well-described by fluid dynamics.
We have seen significant progress in characterizing and quantifying the macroscopic properties of the QGP liquid. 
In the short wavelength limit, the behavior of  asymptotic QGP should be reliably described by the standard perturbative theory.
In contrast, our understanding of QGP at the ''mesoscopic scale''--the scale at which typical wavelength is comparable to Debye length--is rather limited. 
In this least-explored regime, 
characteristic wavelength might be too short for a fluid description of QGP, and is too long for a perturbative treatment. 
In our opinion, 
exploring and understanding the ``mesoscopic physics'' of QGP is one of the key frontiers of the field in the future.

When a jet that is created in a heavy-ion collision passes through the QGP, 
it will deposit energy and momentum to the medium, exciting the medium at both short and long wavelengths.
By studying observables sensitive to jet-induced medium excitations, 
one might be able to probe QGP at the mesoscopic scale and study the QGP evolution as a function of wavelength.

In Ref.~\cite{paper}, 
we investigate the response of a Bjorken-expanding quark-gluon plasma (QGP) to the passage of an energetic parton.
We treat the problem using the linearized relaxation time approximation of the Boltzmann equation.
Although the referred work has no immediate phenomenological goals, 
we wish to demonstrate the intrinsic difference between hydrodynamic and non-hydrodynamic response. 
This can guide future studies to find sensitive probes to the mesoscopic scale QGP in jet-related observables. 
For this purpose, we compare the response functions obtained in Boltzmann theory and in viscous hydrodynamics, 
see Refs~\cite{Hong:2011bd} for related studies.

The evolution of jet-induced medium excitations of energy and momentum is determined by the response function (see more below).
In what follows, 
we will focus on the behavior of the response function itself.  
In this short proceeding, 
we shall briefly discuss selected results from the upcoming publication~\cite{paper}.

%
%

%

\section{The response function and kinetic theory}

Let us consider a small initial disturbance in stress-energy tensor $T^{\mu\nu}$ on top of the background evolution of an expanding QGP.
The (stress-energy tensor) response function describes the evolution of $T^{\mu\nu}$ induced by this initial disturbance. 
One can extract transport coefficients of the medium from the response function in hydrodynamic limit. 
The response function also probes the effective degrees of freedom of the medium at a given scale. 

The response function can be decomposed into a number of independent components. 
In this proceeding, we will focus on one particular component, 
namely, 
the energy-energy response function $G$ that describes energy density $\d \e(\tau,\vx_{\perp})$ induced by an energy disturbance at initial time $\tau_{0}$ and at a point $\vx'_{\perp}$
\begin{eqnarray}
\label{G-def}
\d \e(\tau, \vx_{\perp})
= G(\tau,\tau';\vx_{\perp}-\vx'_{\perp})\, \d\e(\tau',\vx'_{\perp})\, . 
\end{eqnarray}  
For simplicity, we consider the background to be a Bjorken expanding QGP that is spherically symmetric in the transverse plane.
We also limit ourselves to the disturbance in the transverse plane. 
In Eq.~\eqref{G-def}, $\tau=\sqrt{t^2-z^2}$ denotes the proper time and $\vx_{\perp}$ is the transverse coordinate. 
First, we compute $G$ by solving a conformal relaxation time Boltzmann equation in Fourier space, then transform it back to real space. 
We shall use a constant relaxation time $\tau_{R}$.


%

\section{Results}
In this section, we show  $G$ in both Fourier space as a function of wave vector $k$ and in real space as a function of $r=|\vx_{\perp}-\vx'_{\perp}|$ at various snapshots $\Delta\tau=\tau-\tau_{I}$.
Since the background energy density decays as $(\tau/\tau_{I})^{-4/3}$ during the Bjorken expansion, we normalize $G$ by the same factor in all plots shown below. 
We pick $\tau_{I}=\tau_{R}$ in our calculations.

%
%
%
%
%
%

%
%
%
\begin{figure}[t] 
\center
\includegraphics[width=0.95\textwidth]{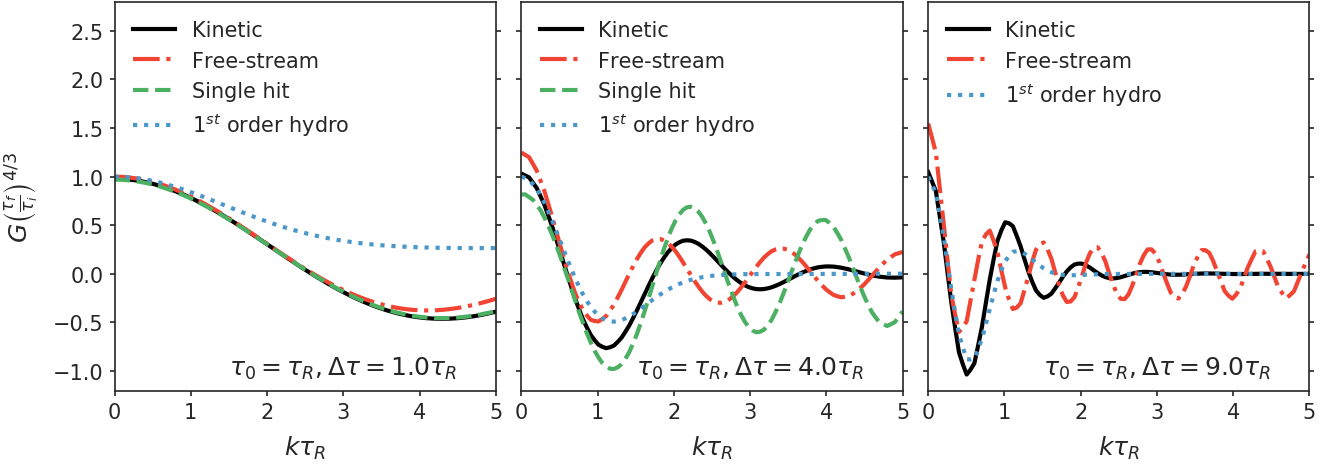}
\caption{
\label{fig:Gttk}
The energy-energy response function in Fourier space plotted as a function of $k\tau_{R}$ at three representative values of $\Delta\tau$.
From left to right, the columns correspond to choosing $\Delta\tau=1,4,9~\tau_{R}$. 
We compare Boltzmann results (black solid curve) with viscous hydrodynamics (blue dotted curve) and with free streaming evolution (red dashed curve).
 }
 \end{figure}
%
%
%

In Fig.~\ref{fig:Gttk}, 
we present $G$ vs $k\tau_{R}$ resulting from solving Boltzmann equation at three representative values of $\Delta \tau=1,4,9$~$\tau_{R}$.
We begin our discussion with $G$ at $\Delta\tau=1~\tau_{R}$ which represents the distribution of energy density shortly after the initial disturbance (see the left column of Fig.~\ref{fig:Gttk}).  
At this stage, the response is dominated by the particle-streaming term in Boltzmann equation that describes the ballistic motion of quasi-particles. 
Such \textit{ballistic response} induces a phase factor $e^{-i \vk\cdot\vv_{\vp} \tau}$ to the perturbed distribution function $\d f$. 
Consequently, 
we observe $G$ oscillates as a function of $k$.
Furthermore, 
we compare Boltzmann results to the free-streaming limit (i.e. neglecting collision integral in the Boltzmann equation), and observe the good agreement.
A better approximation is the ``single-hit'' solution, which includes up-to-one-collision effects to the free-streaming solution.
The single-hit solution well reproduces  Boltzmann results at this short $\Delta\tau/\tau_{R}$. 
In contrast, 
hydrodynamic response  $G_{\hydro}$ at early time is qualitatively different from $G$ even at small $k$. 

Turning to larger values of $\Delta\tau$ as shown in the middle and right column of Fig.~\ref{fig:Gttk}, 
we see the transition from hydrodynamic response to non-hydrodynamic response with increasing $k$. 
This is nicely demonstrated by comparing hydrodynamic results $G_{\hydro}$ with $G$. 
At small wave number $k\tau_{R}\leq 1$, we observe good agreement between $G_{\hydro}$ and $G$ at $\Delta \tau =4~\tau_{R}$ and even more so at $\Delta \tau=9~\tau_{R}$. 
The oscillation in $k$ in this regime arises from the sound propagation.
However, for $k\tau_{R}\geq 1$ where  hydrodynamics approximation breaks down,  $G_{{\rm hydro}}$ is significantly smaller than $G$. 
This is expected since the damping rate of hydrodynamic excitations grows quadratic in $k$, but is less pronounced for quasi-particle excitations.
That said, we emphasize that this does not imply that collisional effects are not important. 
In the contrary, in this large $k$ regime, 
neither free-streaming limit nor single-hit approximation can describe $G$ for $\Delta \tau \geq \tau_{R}$.

%
%
%

%
%
%
\begin{figure}[t] 
\center
\includegraphics[width=0.95\textwidth]{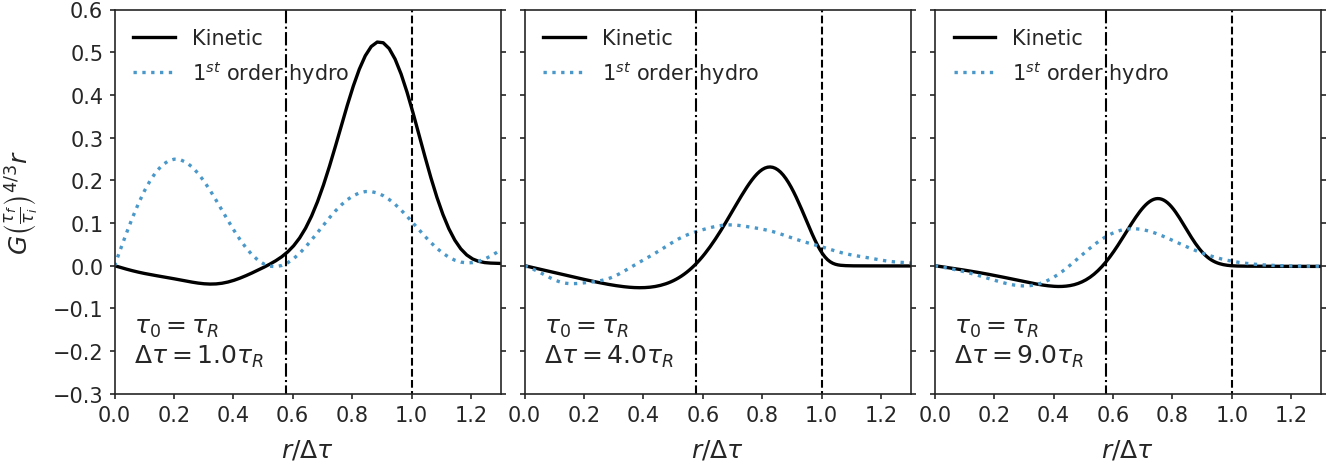}
\caption{
\label{fig:Gr}
The energy-energy response function in real space plotted as a function of $r/\Delta\tau$.  
From the left to the right, the dashed vertical lines show $r/\Delta\tau=c_{s}=0.57$ and $r/\Delta\tau = 1$. 
Since $G\sim 1/r$ at small $r$, we have multiplied $G$ by $r$.
 }
 \end{figure}
%
%
%

In Fig~\ref{fig:Gr}, we present $G$ in real space by (inverse) Fourier transforming results shown in Fig.~\ref{fig:Gttk}. 
It is convenient to plot $G$ as a function of $v\equiv r/\Delta \tau$. 
We observe from Fig.~\ref{fig:Gr} that $G$ features a peak at $v=v_{{\rm eff}}$ for each given $\Delta \tau$. 
We might interpret $v_{{\rm eff}}$ as the effective propagation speed (group velocity) for the energy-momentum disturbance. 
Fig.~\ref{fig:Gr} shows that $v_{\peak}$ approaches the speed of sound $c_{s}$ at late time due to sound propagation and around $1$ at early time since quasi-particle travels at the speed of light. 
However, at an intermediate, say $\Delta \tau=4\tau_{R}$, the deviation from both hydrodynamic and free streaming expectation is significant. 
In this case, the characteristic momentum/frequency is too small for the application of free-streaming approximation, but too large for applying a hydrodynamic description. 
We may view the behavior of the real space response function at such intermediate values of $\Delta \tau$ as representatives of \textit{``mesoscopic response''}. 

We further compare ``mesoscopic response'' with hydrodynamic response by looking at the width of the peak of $G$.
Previously, we have already seen in Fig~\ref{fig:Gttk} that  hydrodynamic response at large $k$ is over-damped.
In Fig.~\ref{fig:Gr}, the width of the peak is broader in hydrodynamic results than that of Boltzmann results at $\Delta\tau=4\tau_{R}$ and $\Delta\tau=9\tau_{R}$ , so evidently the ``mesoscopic response'' is less diffusive than hydrodynamic one.  
At large distances and in the long time limit, 
the disturbance produced by the jet forms a Mach cone and diffusive wake. 
However, 
because of the difference between hydrodynamic and ``mesoscopic`` response in the aspect of both  the propagation and diffusion, the Mach cone and diffusive wake should not be present when ``mescoscpic dyamics'' are significant.

\section{Summary and outlook}

In this proceeding, 
we have studied the response function describing the evolution of energy density perturbation induced by an initial disturbance, using Boltzmann equation under relaxation time approximation. 
In the long time limit,
the response function is well-described by hydrodynamic approximation and in the short time limit, by the ballistic response.
However, at intermediate $\Delta \tau$, 
both hydrodynamic and ballistic description significantly deviate from the  Boltzmann solution.
This is the regime that reflects the physics at the mesoscopic scale. 
In our study, ``mesoscopic response'' is very important at $\Delta\tau=4\tau_{R}$ (and even so at $\Delta\tau=9\tau_{R}$).
Keeping in mind that $\tau_{R}$ is of the order $1$~fm for QGP, the ``mesoscopic response'' can be significant when studying jet-medium interaction from a practical perspective. 

One of the most important qualitative features of our results on ``mesoscopic response'' is that its effective group velocity $v_{\eff}$ for the propagation of energy-momentum disturbance would be greater than sound velocity.
Since this result interpolates the value of $v_{\eff}$ between hydrodynamic limit and free streaming limit, 
it would seem generic for systems with quasi-particles at short distances such as asymptotic free QGP. 
Thus, by studying observables that are sensitive to the propagation speed of medium excitation, 
we might answer whether mesoscopic regime is probed through jet-medium interaction.


We observe viscous hydrodynamic equation can not describe the response function in the ``mesoscopic regime'' in our current model study.
The non-hydrodynamic response in this regime is characterized by a supersonic effective group velocity and a dissipative rate smaller than the viscous damping.
If our goal is to study the mesoscopic scale physics of QGP through jet-medium interaction,
we should implement non-hydrodynamic response in future quantitative modeling.
In Ref.~\cite{paper}, 
we construct a novel model which is similar to M\"uller-Israel-Stewart (MIS) theory, but contains two additional free parameters that respectively control the effective group velocity and dissipation in the non-hydrodynamic regime. 
We demonstrate how to use this model to describe both the non-hydrodynamic and hydrodynamic response in one and the same framework, and discuss how to apply this model to extract the properties of QGP at the ``mesoscopic scale'' through jet-medium interaction, see Ref.~\cite{paper} details.


\textbf{Acknowledgement}-- This work was supported in part by UCB-CCNU Collaboration Grant (WK) and in part by the Strategic Priority Research Program of Chinese Academy of Sciences, Grant No. XDB34000000 (YY).

\end{document}